\documentclass[aps,prr,reprint, onecolumn, superscriptaddress]{revtex4-2}

\usepackage{graphicx}
\usepackage{indentfirst}
\usepackage{amsmath}
\usepackage{mhchem}
\usepackage{pifont}
\usepackage{float}
\usepackage{epstopdf}

\newlength{\myfigwidth}
\begin{document}
%\linenumbers

\title{Slow-light-enhanced Atomic Frequency Comb Quantum Memory in Stoichiometric \ce{EuCl3.6D2O}}

\author{Zongfeng Li}
% \thanks{Z.F. Li and W.T. Xiao contributed equally to this work.}
\thanks{These authors contributed equally to this work.}
\email{zongfeng.li@northwestern.edu}
\altaffiliation[Present address: ]{Department of Electrical and Computer Engineering, Northwestern University, Evanston, IL 60208, USA}
\affiliation{International Quantum Academy, and Shenzhen Branch, Hefei National Laboratory, Shenzhen 518048, China}
\affiliation{Southern University of Science and Technology, Shenzhen 518055, China}

\author{Wanting Xiao}
\thanks{These authors contributed equally to this work.}
% \thanks{Z.F. Li and W.T. Xiao contributed equally to this work.}
\affiliation{International Quantum Academy, and Shenzhen Branch, Hefei National Laboratory, Shenzhen 518048, China}
\affiliation{Southern University of Science and Technology, Shenzhen 518055, China}

\author{Mucheng Guo}
\affiliation{International Quantum Academy, and Shenzhen Branch, Hefei National Laboratory, Shenzhen 518048, China}
\affiliation{Southern University of Science and Technology, Shenzhen 518055, China}

\author{Shuping Liu}
\affiliation{International Quantum Academy, and Shenzhen Branch, Hefei National Laboratory, Shenzhen 518048, China}
\affiliation{Southern University of Science and Technology, Shenzhen 518055, China}

\author{Fudong Wang}
\email{fdwang.phys@foxmail.com}
\affiliation{International Quantum Academy, and Shenzhen Branch, Hefei National Laboratory, Shenzhen 518048, China}
\affiliation{Southern University of Science and Technology, Shenzhen 518055, China}

\author{Manjin Zhong}
\email{manjin.zhong@gmail.com}
\affiliation{International Quantum Academy, and Shenzhen Branch, Hefei National Laboratory, Shenzhen 518048, China}
\affiliation{Southern University of Science and Technology, Shenzhen 518055, China}

\date{\today}

\begin{abstract}

Rare-earth-doped crystals are promising candidates for quantum storage, yet their performance in free-space configurations is fundamentally restricted by low optical depth. Here, we demonstrate high-efficiency quantum storage in a stoichiometric \ce{EuCl3.6D2O} crystal, which intrinsically provides high optical density without the complexity of cavity implementation. 
We show that in this high-density regime, the system exhibits significant slow-light-like effects, including dispersion-induced echo delays and finesse-dependent echo intensity modulation. 
We develop a unified theoretical framework showing how absorption and dispersion work in concert to mediate echo generation. 
We achieve storage efficiencies of 42.9\% for classical light and 34.4\% for weak coherent pulses, alongside 90\% efficiency for slow-light storage. 
These findings validate \ce{EuCl3.6D2O} as a robust platform, establishing a viable pathway for scalable solid-state quantum memory.

\end{abstract}

%\keywords{}

\maketitle


\section{Introduction}

Quantum networks require reliable long-range entanglement, typically achieved via quantum repeater protocols \cite{briegel1998quantum, simon2007quantum, sangouard2011quantum}. Efficient quantum memories are essential components of these repeaters. Rare-earth (RE) ion-doped crystals have emerged as leading candidates for quantum storage due to their exceptionally long coherence times \cite{zhong2015optically, 1hour, PRXQuantum.6.010302}, multimode capacity \cite{afzelius2009multimode, wei2024quantum, yang2018multiplexed}, and integration potential \cite{zhong2017nanophotonic, zhong2019emerging}. Typically, these systems operate at low ion concentrations to mitigate decoherence and line broadening caused by ion-ion interactions and lattice distortion \cite{equall1995homogeneous}. However, this dilute approach inherently limits the optical depth (OD), thereby constraining storage efficiency \cite{afzelius2010impedance, sabooni2013efficient}.

To overcome the limitations of low optical depth in dilute RE-doped crystals, several strategies have been pursued. While cavity enhancement and multipass configurations can effectively increase interaction length, they often introduce significant technical complexities \cite{davidson2020improved, schraft2016stopped}. In contrast, traditional free-space atomic frequency comb (AFC) storage schemes are inherently limited by low ion density, with experimental efficiencies typically plateauing around 40\% \cite{Maring2018thesis, horvath2021noise, ortu2022multimode}. Consequently, there is a strong motivation for a medium that retains the simplicity of free-space configurations while providing the high optical depth necessary for high-efficiency storage. Stoichiometric RE crystals offer an intrinsic route to this goal due to their orders-of-magnitude higher ion density and exceptionally high optical depth (OD) \cite{rose2013, ahlefeldt2016ultranarrow}.
%, though maintaining long coherence times in such dense systems remains a challenge \cite{schlittenhardt2024spectral}.

In this highly dense regime, the interplay between absorption and dispersion manifests with distinct experimental consequences that are often negligible in low-density systems. Specifically, the strong dispersive effects in such dense media enable new control over storage dynamics, such as the manipulation of echo timing and intensity, offering a practical pathway to enhance memory performance beyond the limitations of standard absorption-based models.

In this work, we investigate the potential of a home-grown \ce{EuCl3.6D2O} crystal for high-efficiency quantum storage. We demonstrate high AFC storage efficiencies of 42.9\% for classical light and 34.4\% for weak coherent pulses without cavity enhancement. Beyond experimental performance, we show that the storage dynamics in this dense system can be described as a coherent superposition of slow-light modes. We provide a unified theoretical framework that accounts for both absorption and dispersion to reveal how these dynamics can be exploited to modulate echo timing and extend storage time. Our findings validate \ce{EuCl3.6D2O} as a robust platform for quantum storage and offer new insights into controlling light propagation in highly absorptive rare-earth media.

\section{Efficient AFC and spectral-hole memory in \ce{EuCl3.6D2O}}

\begin{figure}[htb]
\centering\includegraphics[width=0.8\linewidth]{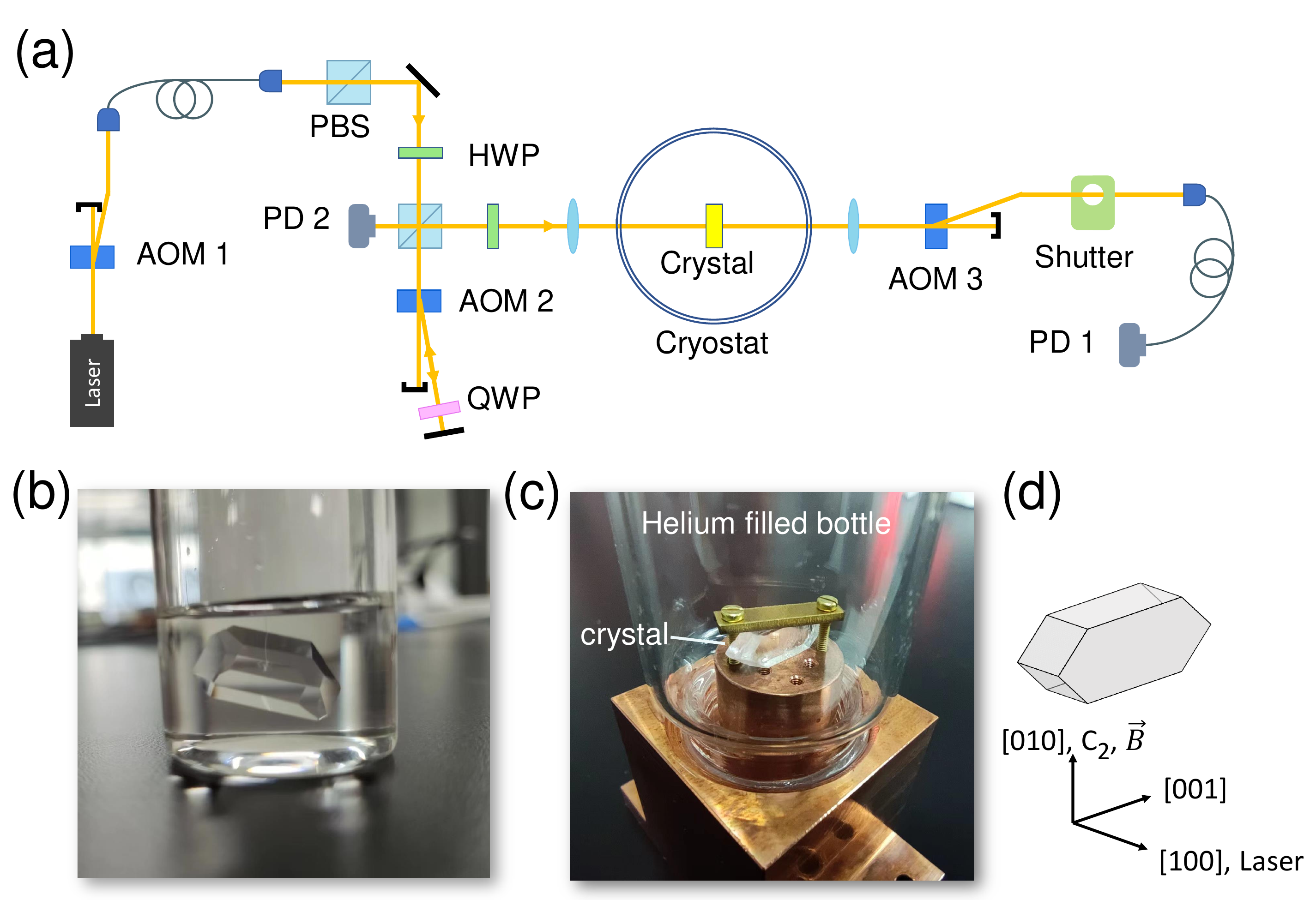}
\caption{Schematic diagram of the optical experimental setup and crystal orientation. 
(\textbf{a}) AOM: acousto-optic modulator, HWP: half wave plate, QWP: quarter wave plate, PD: photodiode. (\textbf{b}) The \ce{EuCl3.6D2O} single crystal grown in a heavy water solution. (\textbf{c}) The crystal is mounted in a helium filled bottle and on a copper base. (\textbf{d}) Crystal cleaved along the (100) plane. }
\label{fig_setup}
\end{figure}

We grew a \ce{EuCl3.6D2O} crystal from a \ce{EuCl3} heavy-water solution. The \ce{EuCl3.6D2O} was cleaved along the (100) plane \cite{ahlefeldt_minimizing_2013}, mounted in a dilution refrigerator, and cooled down to 100 mK, as illustrated in Fig.~\ref{fig_setup}.
Hole-burning pulses were generated using two acousto-optic modulators (AOMs); further experimental details are provided in Appendix~\ref{setup}. 
We prepared the AFC by burning periodic frequency spectral holes on the low-frequency side of the $^7\mathrm{F}_0 - {}^5\mathrm{D}_0$ transition, targeting $\mathrm{Eu}^{3+}$ ions fully surrounded by D rather than H to reduce phonon-induced nonradiative decay, as detailed in Appendix~\ref{spect} \cite{rose2013}.

\begin{figure}[htb]
\centering\includegraphics[width=0.8\linewidth]{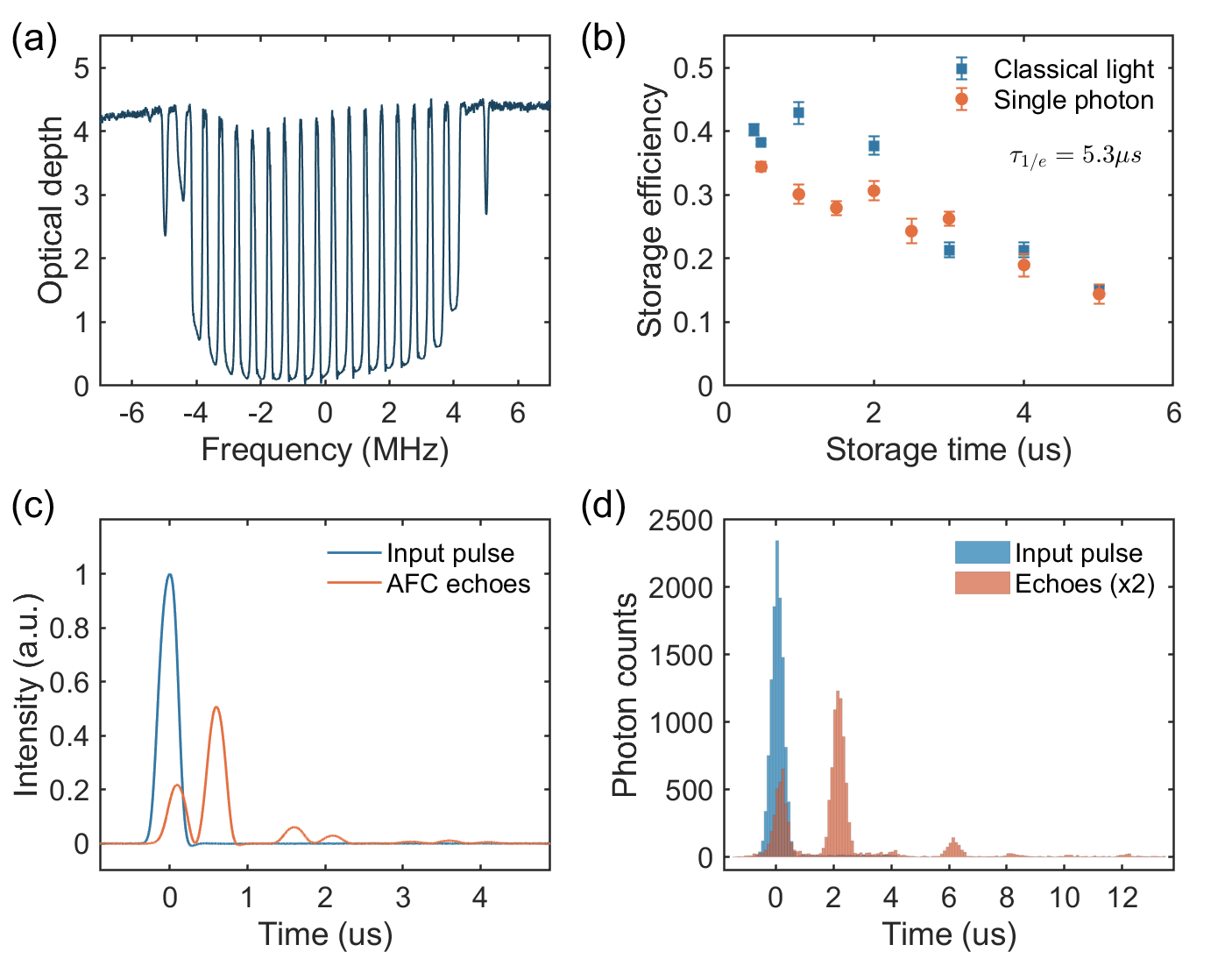}
\caption{Efficient AFC storage in \ce{EuCl3.6D2O}.
(\textbf{a}) An 8-MHz-wide AFC spectrum with a storage time of 2~$\mu\mathrm{s}$ is burned by two parallel 4-MHz-wide comb burning pulses centered at -2~MHz and 2~MHz, to avoid over-pumping at the center and under-pumping at edges. (\textbf{b}) The storage efficiency of classical light and single photons at various storage times. An exponential fit to both datasets gives a $1/e$ storage time of 5.3~$\mu\mathrm{s}$. (\textbf{c}) Average trace (1000 runs) of a classical light storage of 0.5~$\mu\mathrm{s}$ and (\textbf{d}) 6000 times integration of weak coherent light storage of 2~$\mu\mathrm{s}$. }
\label{fig_AFC}
\end{figure}

Figure~\ref{fig_AFC}(a) displays a typical prepared comb-like absorption structure. Such a periodic profile allows an incoming light pulse to be stored and later retrieved as a photon echo \cite{afzelius2009multimode}.
The directly measured optical depth, defined as $-\ln{(I_{\mathrm{out}}/I_{\mathrm{in}})}$, is limited to approximately 4.5 by the detector's dynamic range; however, the actual OD is estimated at $\sim 25$ (see Fig.~\ref{Delay}). We achieved optimal storage efficiency for classical light by tuning the AFC finesse and pump pulse repetitions. These optimized parameters were subsequently applied to weak coherent pulse storage. As shown in Fig.~\ref{fig_AFC}(b), the time-dependent storage efficiency yields a $1/e$ storage lifetime of 5.3~$\mu\mathrm{s}$. We achieved peak storage efficiencies of 42.9 $\pm$ 1.7\% (at 1~$\mu\mathrm{s}$ storage time) for classical light and 34.4 $\pm$ 0.8\% (at 0.5~$\mu\mathrm{s}$) for weak coherent pulses. To our knowledge, this represents the highest AFC storage efficiency reported for Eu-based crystals without cavity or waveguide enhancement. Typical AFC echoes for classical and weak coherent light are shown in Fig.~\ref{fig_AFC}(c,d).

\begin{figure}[htb]
    \centering
    \includegraphics[width=0.7\linewidth]{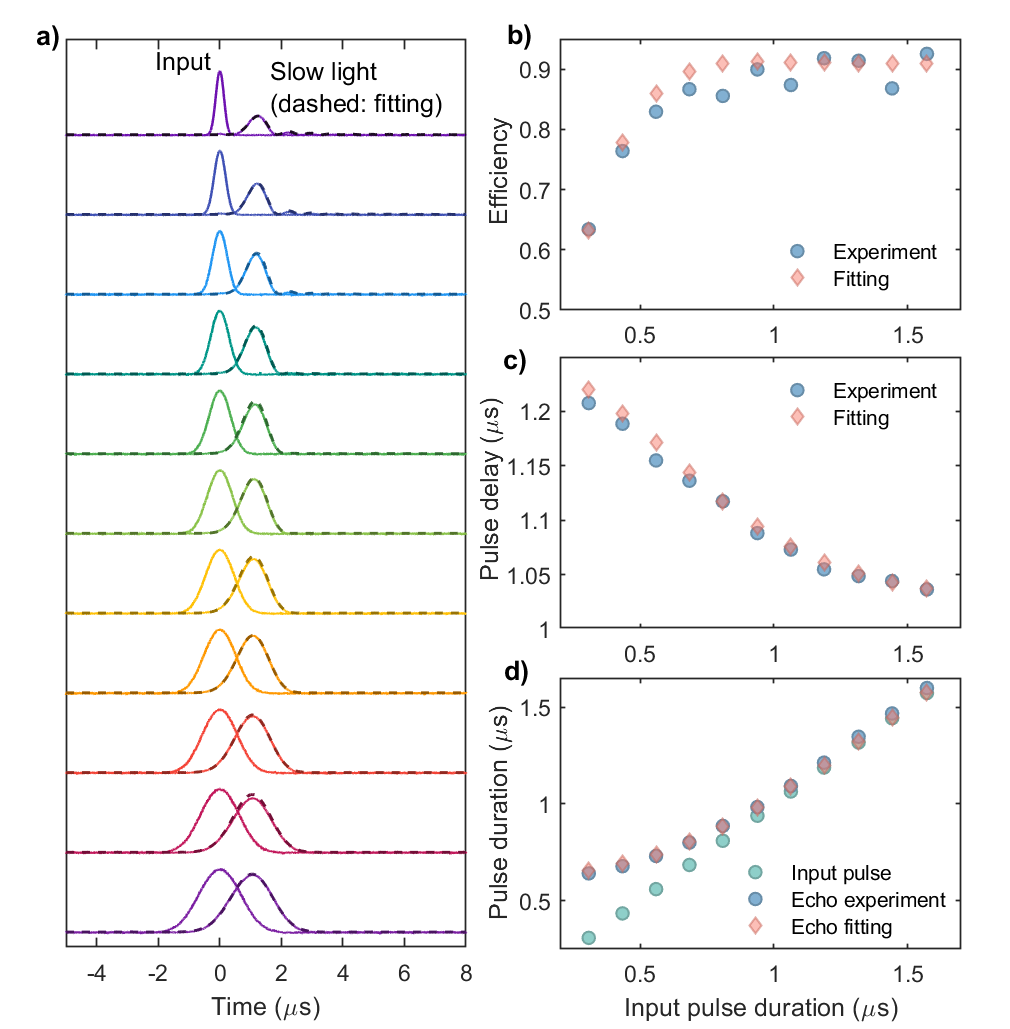}
    \caption{Spectral hole memory (slow-light). 
    (\textbf{a}) A pulse with varying duration(bandwidth) sent into a 1-MHz-wide spectral hole. (\textbf{b}), (\textbf{c}) and (\textbf{d}) The comparison of experiment and fitting results of efficiency (pulse area), delay time and pulse duration, respectively. }
    \label{fig_sl}
\end{figure}

Spectral-hole memory (slow-light) utilizes a refractive index gradient within a persistent spectral hole to reduce the group velocity of a pulse with a matched bandwidth \cite{Spectral_hole}. Figure~\ref{fig_sl} shows the results for input pulses of varying duration and bandwidth injected into a 1-MHz-wide spectral hole, achieving an efficiency exceeding 90\% with a 1~$\mu\mathrm{s}$ delay. Pulses with shorter durations (and larger bandwidths) exhibit lower efficiency because only the spectral component within the hole bandwidth is slowed down (Fig.~\ref{fig_sl}(b,d)). The refractive index gradient is steeper near the hole edge than at the center (Fig.~\ref{Delay}(d)), which explains why narrowband inputs experience a slightly shorter delay (Fig.~\ref{fig_sl}(c)). Our model assumes an ideal square hole, as detailed in the following section; fitting yields an optical depth of 18 and a bandwidth of 1.8~MHz.

\section{Slow-light effect in AFC and theoretical explanation}

\subsection{Modulation of echo intensities via AFC finesse}

% Phenomenon 2
\begin{figure}[htb]
\centering\includegraphics[width=0.8\linewidth]{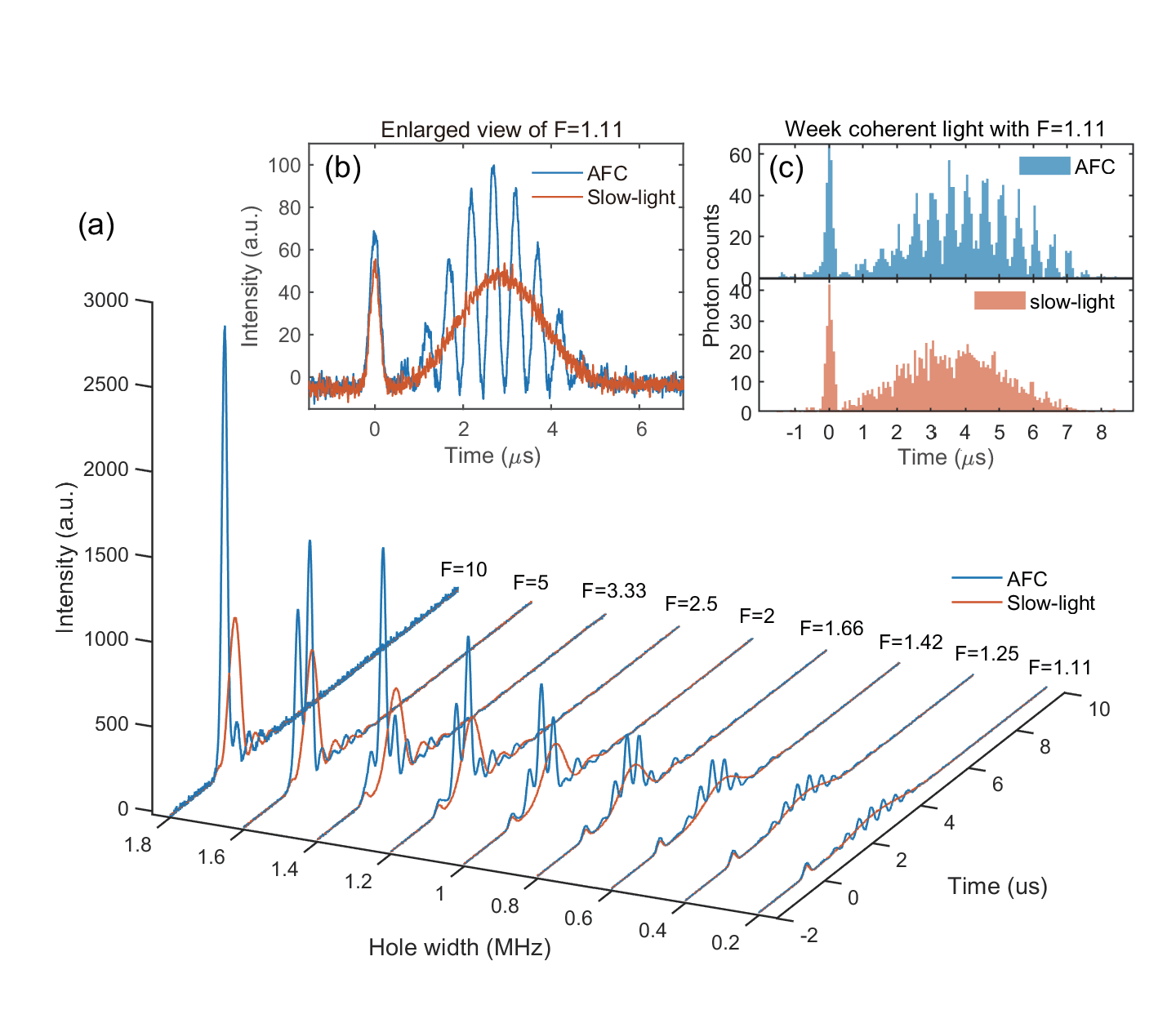}
\caption{(\textbf{a}) AFC memory (blue, composed of 6 spectral holes and $\Delta$= 2~MHz) versus one-hole slow-light memory (orange, identical hole width to the AFC) of an identical input field, under a varying hole width (AFC finesse $F$). (\textbf{b}) Enlarged view of the $F = 1.11$ curve in (\textbf{a}). (\textbf{c}) AFC and slow-light memory traces for weak coherent light, with a hole width of 0.2~MHz.}
\label{interfer}
\end{figure}

% 2 phenomenon
%Equipped with the theoretical framework, we now analyze two significant and unexpected slow-light-like effects observed in our AFC storage experiments.
We now analyze two significant and unexpected slow-light-like effects observed in our AFC storage experiments. 
The theoretical framework for the corresponding explanations, including the intensity of echoes at each order and the time-domain output field is given in Appendices~\ref{analytical},~\ref{numerical}.
The first notable observation is that the intensities of higher order AFC echoes exceed those of lower-order echoes at low AFC finesse, as shown by the blue trace in Fig.~\ref{interfer}. 
This contrasts with conventional AFC behavior, where the amplitudes of higher-order echoes diminish quickly. 

For comparison, we performed a slow-light memory experiment, in which a single square spectral hole with the same width as the AFC holes is burned (i.e., an AFC with only one hole rather than a comb).
The result is shown in Fig.~\ref{interfer}.
As the hole width (and AFC finesse) increases, the group delay, which is proportional to the dispersion $\frac{\partial n}{\partial \omega}$, decreases, resulting in the energy of the AFC echoes concentrating in the lower-order echoes.
The temporal coincidence of the echoes in two protocols implies that AFC can be considered as a coherent superposition of multiple slow-light memory modes. 

\begin{figure}[htb]
\centering\includegraphics[width=0.8\linewidth]{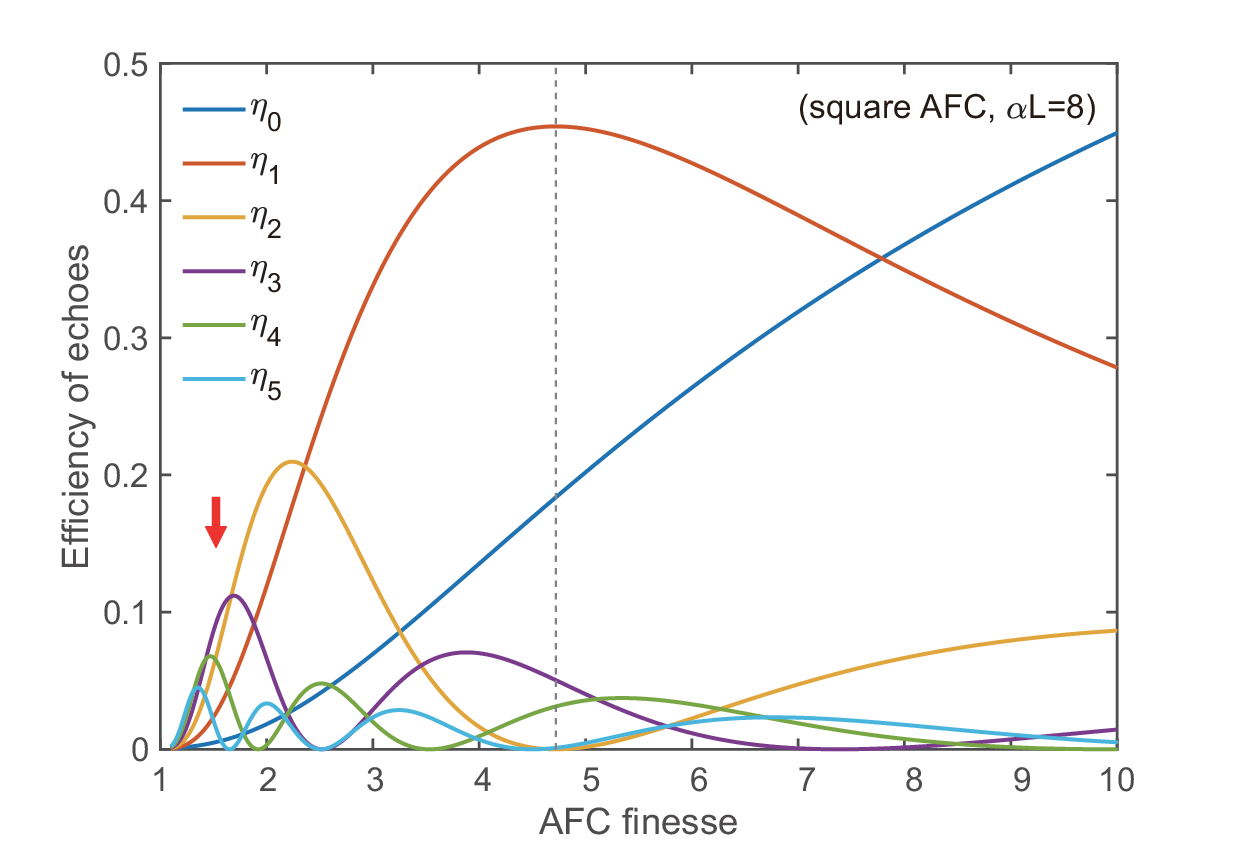}
\caption{Efficiency of the transmitted input ($\eta_0$) and the first 5 echoes ($\eta_{1-5}$) versus AFC finesse. Dashed vertical line: optimal finesse for the 1st echo. Arrow: region where higher-order echoes greater than lower-order echoes.}
\label{fig5}
\end{figure}

% 2, theory and explanation
This phenomenon can be explained by the analytical solutions of the finesse-dependent echo efficiencies ($\eta_n = |a_n(L)|^2$, given by Eqs.~\ref{e15}) and is visualized in Fig.~\ref{fig5}.
At the low finesse region (red arrow), the higher-order echoes exhibit greater efficiencies than lower-order echoes. 
Moreover, because the storage in a two-level ensemble is a linear process, AFC is actually a coherent superposition of multiple slow-light modes, if not transferring the population to a third level.
When the hole width is narrower, group delay is longer and energy concentrates to higher order AFC echoes.
In addition, at the optimal finesse (dashed line), the 2nd and 5th echoes are suppressed, in agreement with the experimental results in Fig.~\ref{fig_AFC}.

\subsection{Dispersion-induced storage time extension}

% Phenomenon 1
\begin{figure}
\centering\includegraphics[width=0.8\linewidth]{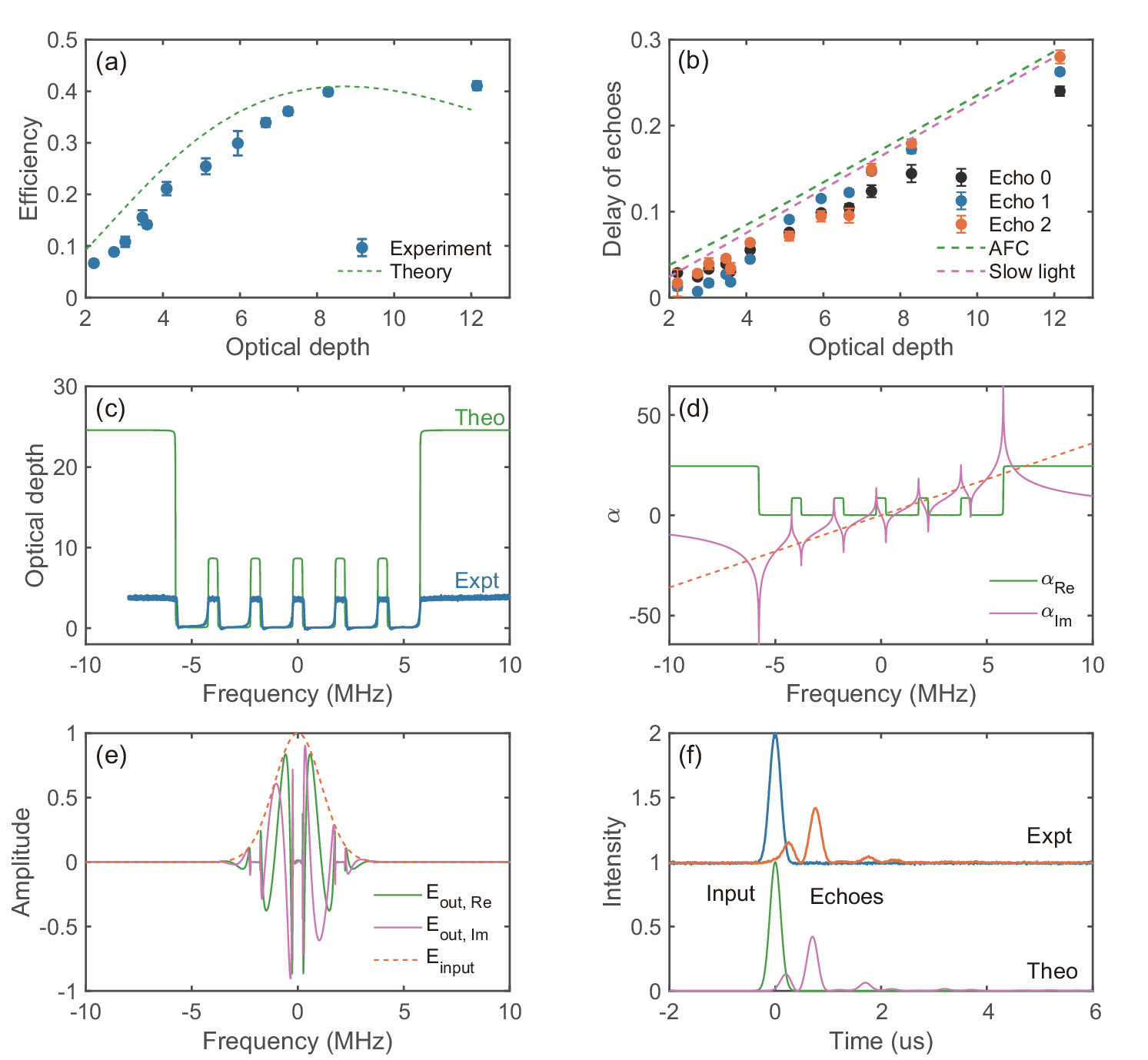}
\caption{
Dependence of AFC echo delay on OD. 
AFC Storage efficiency (\textbf{a}) and echo delay compared to their expected timing (\textbf{b}) at different optical depths (spectral positions). Dashed lines: the corresponding theoretical fits. 
(\textbf{c}) The measured and theoretical absorption spectrum. (\textbf{d}) The real and imaginary parts of complex absorption coefficients of AFC. The dashed orange line reveals the overall dispersion which causes the echo delays.
The output field in the frequency domain (\textbf{e}) and in time domain(\textbf{f}). }
\label{Delay}
\end{figure}
% =========== (a，b) change x axis values ==========
% add a arrow in (a)

% 1， Phenomenon
The second notable observation is that all AFC echoes, including the transmitted input pulse (0th-order echo), are delayed compared to their expected arrival times, as can be observed in Fig.~\ref{fig_AFC}(c, d) and Fig.~\ref{Delay}(f). 
We investigated this unexpected delay by changing the OD (i.e., by tuning the laser frequency) while keeping the AFC parameters constant (comb period $\Delta=2$~MHz, finesse $F=4.4$).
The observed echo delay reached up to 0.25~$\mu\mathrm{s}$, which corresponds to an extension of the storage time by 50\%.

The efficiency and echo delay are shown in Fig.~\ref{Delay}(a,b). 
The horizontal axis represents the OD of AFC peaks, which is calculated from laser frequency by using the fitted absorption curve in Fig.~\ref{spect}(a).
The experimental result and theoretical analysis of the penultimate experimental data point in Fig.~\ref{Delay}(a,b) are shown in (c-f).

% 1， theory
This delayed echo can be accurately calculated by the numerical method, in which the output field intensity vs. time was obtained (Fig.~\ref{Delay}(f)). 
We attribute this unexpected delay to the entire AFC acting as a single wide hole, by averaging out the fine comb structure.
This wide hole (blurred AFC) has a width of $N\Delta$, depth of $\alpha_M(1-1/F)$, where N is the comb number.
The calculated results from both the original and blurred AFCs (Fig.~\ref{Delay}(b)), as well as the group delay formula for a rectangular spectral hole, $\tau_g = \frac{\alpha_M(1-1/F)}{\pi^2 N\Delta}$, all yield the same result.
This indicates the overall dispersion span of the whole AFC range is the reason for the delay of echoes (dashed line in Fig.~\ref{Delay}(d)).

The agreement of those calculations and experiments is based on the assumption that AFC peaks are lower than the original absorption (unpumped outer-side spectra), as shown in Fig.~\ref{Delay}(c). 
This is a reasonable assumption because a low AFC ground OD is more important than a high peak OD in maximizing the storage efficiency. Thus the AFCs are always over-pumped during preparation and AFC peaks are etched by the power broadening.
Furthermore, the experimental delay would be $\sim 3$ times greater than the calculated result if the original absorption were as high as the AFC peaks.
The original absorption is determined by fitting the data in Fig.~\ref{Delay}(a,b).

% 2, verfy

\begin{figure}[htb]
    \centering
    \includegraphics[width=0.6\linewidth]{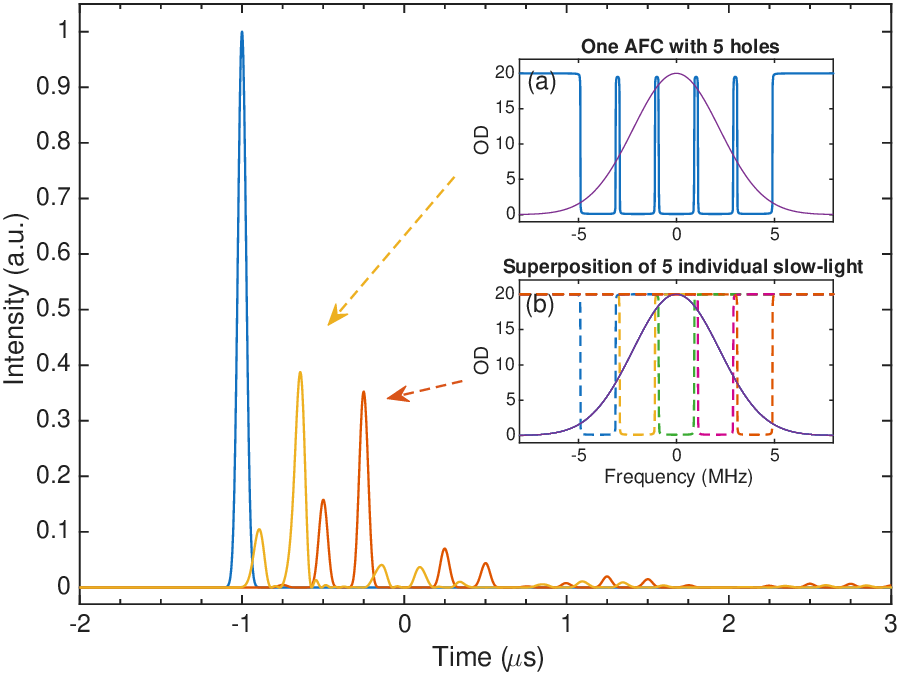}
    \caption{Theoretical AFC echoes (yellow) and the coherent superposition of the light fields of 5 independent slow lights (orange). Inset: (\textbf{a}) the spectrum of a single AFC and (\textbf{b}) 5 individual single-holes; the Gaussian curve is the input pulse.}
    \label{sl-superposition}
\end{figure}

To verify that the AFC echo is the result of a coherent superposition of a set of slow-light modes, we show numerical results of these two cases in Fig.~\ref{sl-superposition}.
The yellow curve shows the echoes of a 5-hole-AFC with storage time of 0.5~$\mu\mathrm{s}$, OD of 20 and finesse of 10, while the orange curve indicates the echoes which are the coherent superposition of 5 individual single-hole slow-light memories (i.e., summing the 5 individual output field amplitudes in time or frequency domain).
The AFC echoes are delayed 0.21~$\mu\mathrm{s}$ by the slow-light effect (0.71~$\mu\mathrm{s}$ in total), while the slow-light superposition echoes align with the integer multiples of storage time, but later (1~$\mu\mathrm{s}$ and 1.5~$\mu\mathrm{s}$). 
This is because the single-hole slow-light does not introduce an overall dispersion in the photon bandwidth, but has a stronger dispersion in the single-hole area compared to the AFC case. 
The similarity of the two echo sequences verifies the superposition explanation of AFC memory.

\subsection{Absorption-dispersion interplay in AFC memory}

Here we discuss the contributions of the absorption and dispersion in AFC memory.
The real and imaginary parts of the complex absorption coefficient $\tilde{\alpha}(\omega)$ are connected by the Kramers-Kronig relation (convolution form),
\begin{equation}\label{eq3}
\alpha_{\mathrm{Im}}(\omega) = -\frac{1}{\pi}\cdot\frac{1}{\omega}*\alpha_{\mathrm{Re}}(\omega).
\end{equation}
By utilizing the Fourier transform relation $\mathcal{F}[\frac{1}{\omega}] = i\pi\cdot \mathrm{sign}(t)$, the simplified field evolution Eq.~(\ref{e7}) can be written as
\begin{equation}\label{eq4}
\partial_z E(z,t) = \frac{\alpha_M}{2\sqrt{2\pi}} \cdot[(1+ \mathrm{sign}(t))\cdot \mathcal{F}[g(\omega)]]*E(z,t),
\end{equation}
where $g(\omega)$ is the normalized ion distribution in the frequency domain (analogous to the absorption spectrum), and the terms $1$ and $\mathrm{sign}(t)$ come from the absorption and dispersion, respectively, and together lead to the Heaviside step function in Eq.~(\ref{e7}).

In AFC memory, frequency-domain periodic distributions of atoms manifest as a series of discrete Dirac $\delta$ functions in the time domain (Eq.~(\ref{e8})). 
These $\delta$ functions in a convolution work as the time delay operators (Eq.~(\ref{e12})).
Dynamic equations of each echo (Eq.~(\ref{e13})) reveal that the $n$th echo is stimulated by applying the delay operators to the set of earlier (0th to $(n-1)$th) echoes.
Recall that the $1$ and the $\mathrm{sign}(t)$ terms represent the contributions of absorption and dispersion, respectively. Therefore, one can conclude:

\begin{enumerate}
    \item An echo will generate the subsequent echoes due to the \textit{equal} contributions of absorption and dispersion.
    \item An echo does not generate the preceding echoes due to the \textit{equal but opposite} contributions of absorption and dispersion, satisfying causality.
    \item The echo by itself is affected only by absorption during propagation.
\end{enumerate}

Since all the echoes are stimulated by earlier echoes and ultimately the input pulse, one can conclude that the absorption and dispersion play equal roles in the generating of higher-order AFC echoes.

It might be argued that it is a general conclusion for all absorption-based quantum storage schemes, as Eq.~(\ref{eq4}) is a direct consequence of the Kramers-Kronig relations and is independent of the specific absorption profile.
However, the key lies in the experience of the incident light field. For instance, in slow-light or EIT storage, the incident photon bandwidth is much smaller than the linewidth of the transparent spectral window. Therefore, the real part of the absorption coefficient vanishes in the evolution equation\cite{Spectral_hole}, leading to a dispersion-driven storage protocol. In contrast, in photon-echo type storage, such as controlled reversible inhomogeneous broadening (CRIB) \cite{kraus2006quantum}, gradient echo memory (GEM) \cite{hedges2010efficient, hosseini2011high} and the revival of silenced echo (ROSE) protocol \cite{damon2011revival}, the absorption spectrum is uniform within the photon bandwidth, exhibiting negligible dispersion. This uniformity results in an absorption-driven storage protocol. 
In contrast, AFC operates in a regime where the probe spans multiple absorption peaks and transparency windows, unlike EIT or CRIB, so both mechanisms are 
non-negligible. 
This distinguishes AFC as a unique protocol where absorption and dispersion contribute equally.

Although absorption and dispersion provide equivalent driving forces from a dynamical perspective, in high-efficiency AFC storage, as concluded by Bonarota et al. \cite{Bonarota_2012}, most of the energy of the field resides in the crystal as slow light, rather than as atomic excitations. This also agrees with the statement that AFC is a coherent superposition of multiple slow-light modes.

\section{Discussion}

To fully exploit the high optical depth of \ce{EuCl3.6D2O} for quantum storage, achieving long coherence times is essential. Analysis of the Eu$^{3+}$ Hamiltonian reveals a significantly lower magnetic sensitivity, and consequently longer coherence times, under zero first-order Zeeman (ZEFOZ) conditions \cite{longdell_characterization_2006, ahlefeldt_minimizing_2013, PRXQuantum.6.010302}.
Our measurements confirm that \ce{EuCl3.6D2O} exhibits optical coherence times up to 1 ms, spectral hole lifetimes exceeding 20 minutes (Appendix~\ref{spect}), and low background absorption after spectral tailoring. These findings underscore the potential of \ce{EuCl3.6D2O} crystals for high-efficiency, long-storage-time quantum memory.

The current storage efficiency is primarily constrained by the 92\% deuterium concentration. Based on our estimates, increasing the D concentration to 99.5\% could yield absorption coefficients of approximately 300 cm$^{-1}$ (Appendix~\ref{spect}). Such high optical depth, combined with the excellent spectral tailoring capabilities of \ce{EuCl3.6D2O}, makes it a strong candidate for storage protocols with theoretical efficiencies approaching 100\%, such as Gradient Echo Memory (GEM) or backward-echo configurations. Regarding on-demand spin-wave quantum memory, the long hyperfine lifetime facilitates the initialization of Eu$^{3+}$ ions into specific spin states \cite{lauritzen2012spectroscopic, zhu2020coherent}.

For single-hole slow-light memory, reducing the hole width to 500 kHz with an optical depth (OD) of 60 could extend delay times to 12~$\mu\mathrm{s}$. 
This capability enables high-efficiency stopped-light protocols, including EIT memory \cite{chen2013coherent, schraft2016stopped} and spin-wave slow-light memory \cite{Spectral_hole}. 
However, implementing high-quality population transfer ($\pi$) pulses in high-concentration systems presents a challenge, as strong absorption necessitates higher control field energies. 
In this work, we demonstrate that incorporating an AFC into a highly absorptive background extends photon residence time by 0.25~$\mu\mathrm{s}$, with potential extensions up to 1.25~$\mu\mathrm{s}$ at higher OD (60) and narrower AFC widths (5~MHz). 
This slow-light-AFC hybrid scheme effectively increases the photon-crystal interaction time, thereby facilitating the extension of $\pi$-pulse durations for on-demand spin-wave memories. While this may impose limits on time-domain multimode capacity, spatial or frequency multiplexing remains fully compatible \cite{pu2017experimental, yang2018multiplexed, li2025efficient}.

Finally, the efficiency-finesse relation established in Fig.~\ref{fig5} offers a new degree of freedom for modulating AFC echoes, such as creating a time-domain atomic beam splitter by tuning the finesse to equalize the first and second echoes. The observed long coherence time and hyperfine lifetime also validate \ce{EuCl3.6D2O} as a competitive material for quantum computing and microwave-to-optical transduction. Specifically, addressing host ions adjacent to dopants \cite{ahlefeldt2013method} could enable two-qubit C-NOT gates via ion-ion interactions, supporting small-scale quantum computing within repeater nodes \cite{ahlefeldt2020quantum}. Furthermore, collective spin excitations (magnons) in stoichiometric RE crystals provide a platform for highly efficient microwave-to-optical conversion—a key technology for interfacing microwave qubits, such as those in superconducting architectures, with optical networks \cite{everts2019microwave, han2021microwave}.

\section{Conclusion}
We have demonstrated high-efficiency quantum storage in a stoichiometric \ce{EuCl3.6D2O} crystal, achieving efficiencies of 42.9\% for classical light and 34.4\% for weak coherent pulses using the AFC protocol. 
Our findings reveal that AFC storage functions as a coherent superposition of slow-light modes, and the cumulative dispersion across the whole AFC naturally induces storage time extensions. 
By leveraging this slow-light-enhanced interaction, we effectively increase the photon residence time within the crystal. 
Our analysis demonstrates that absorption and dispersion contribute equally but differently to the echo generation, providing an accurate theoretical method for static spectrum quantum storage protocol. 
These results also validate stoichiometric \ce{EuCl3.6D2O} as a promising candidate for efficient quantum memories.

\begin{acknowledgments}
This work was supported by the Quantum Science and Technology -National Science and Technology Major Project (No. 2021ZD0301204), the National Natural Science Foundation of China (Grant No. 11904159, 12004168 and 12304454), Guangdong Innovative and Entrepreneurial Research Team Program (Grant No. 2019ZT08X324), Guangdong Basic and Applied Basic Research Foundation (Grant No. 2021A1515110191), and the Key-Area Research and Development Program of Guangdong Province (Grant No. 2018B030326001), the Science, Technology and Innovation Commission of Shenzhen Municipality (KQTD20210811090049034).
\end{acknowledgments}

\appendix
\section{Crystal growth and experiment setup}
\label{setup}

The crystal used in these experiments is a deuterated \ce{EuCl3.6D2O} crystal. 
Deuteration, the substitution of hydrogen (H) with deuterium (D), is crucial for extending the optical lifetime and coherence time in rare-earth hydrate crystals by reducing non-radiative transitions  \cite{heller1966formation, rose2013JoL}.
To prepare the crystal, 10 g of \ce{EuCl3.6H2O} was repeatedly dissolved in 5 g of 99.96\% \ce{D2O} and evaporated using a rotary evaporator \cite{rose2013JoL}. 
A seed crystal was then suspended in 10 mL of a saturated deuterated solution and cooled from 30$^{\circ}$C to 20$^{\circ}$C over two weeks \cite{myPLA} to grow the \ce{EuCl3.6D2O} single crystal (Fig.~\ref{fig_setup}(b)).
The crystal was cleaved along the (100) plane \cite{ahlefeldt_minimizing_2013}, with a 2-mm-thickness along the direction of light propagation ([100]). 
The crystal was then mounted in a helium-filled, hermetically sealed glass vial to isolate it from the atmosphere or vacuum (Fig.~\ref{fig_setup}(c)).
The crystal was cooled to 100 mK in a dilution refrigerator (BlueFors).  
The $^7\mathrm{F}_0 \rightarrow {}^5\mathrm{D}_0$ transition of Eu$^{3+}$ exhibits maximum (minimum) absorption when the light polarization is parallel (perpendicular) to the [010] direction (C2 axis).%, as shown in Fig.~\ref{fig_spectrum}(d).

The experimental setup is shown in Fig.~\ref{fig_setup}(a).  
Single-pass (AOM1) and double-pass (AOM2) acousto-optic modulators generate pump and input pulses (bright or single-photon level).  
A photodiode monitors power drift. 
After the crystal, the beam is gated by a single-pass AOM and a mechanical shutter to protect the detector from the bright pulse.  Detection is performed using a photodiode for bright pulses, a photomultiplier tube for fluorescence, or a single-photon detector for single photons.  
AFCs are prepared by directly burning several square holes with separation $\Delta$ by complex hyperbolic secant (CHS) pulses \cite{silver1985selective, rippe2005experimental, jobez2016towards}. 
When the storage time exceeds 2~$\mu\mathrm{s}$, the square holes are burned by parallel CHS pulses. For shorter storage times, each square hole is burned individually due to the larger total bandwidth.
The pump pulses repeat 200-1500 times, with an interval of 5 ms, depending on different absorption depth.
A Gaussian pulse enters the crystal 10~ms after the AFC preparation, and is retrieved after the storage time $1/\Delta$.
During the weak coherent light storage, AOM1 and AOM2 attenuate the input pulse such that the average number of photons per pulse is approximately 1.

\section{Spectroscopic characterization}
\label{spect}

\begin{figure}[htb]
\centering\includegraphics[width=12cm]{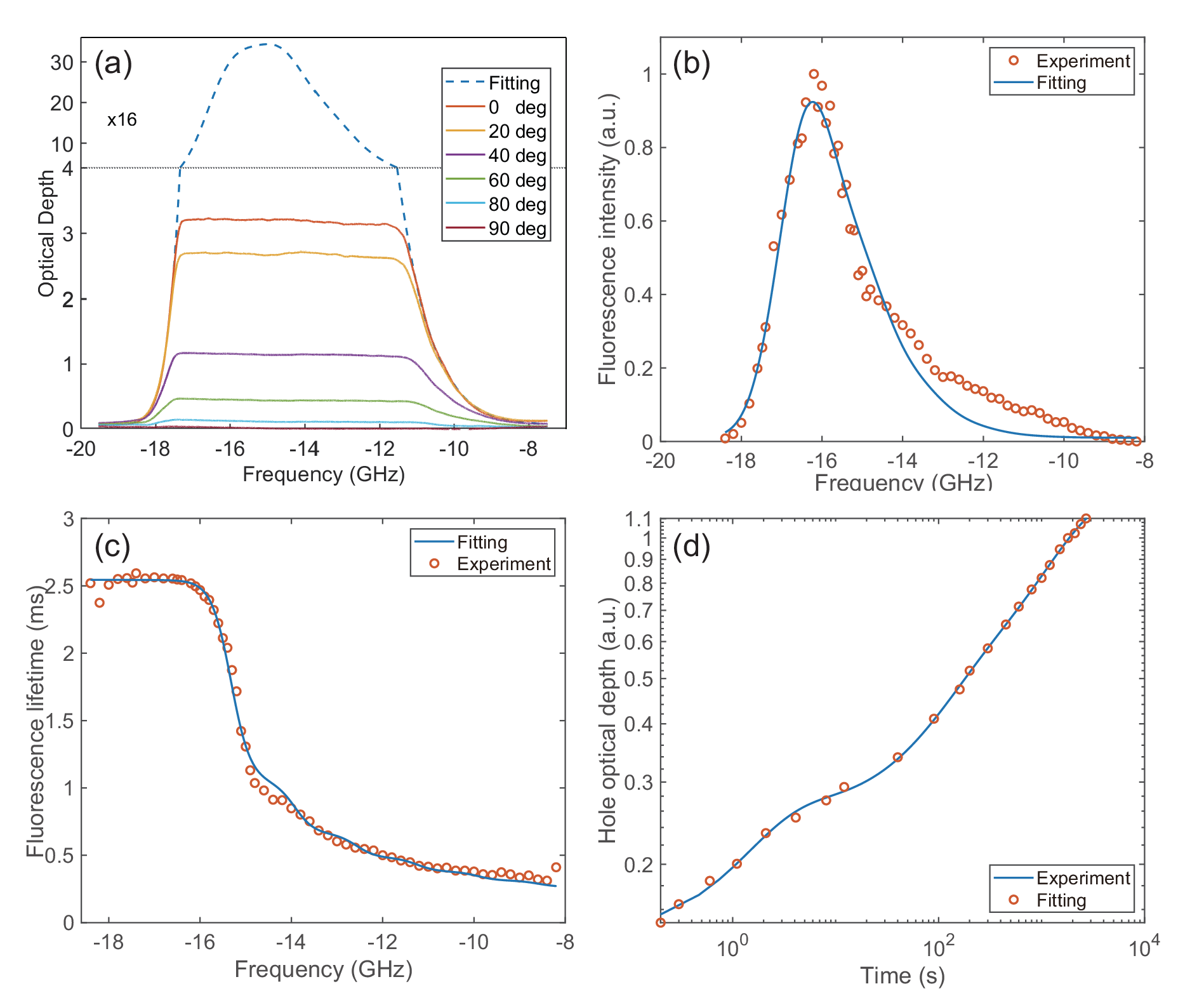}
\caption{Spectroscopic characterization of \ce{EuCl3.6H2O}.  (\textbf{a}) Absorption spectra of \ce{EuCl3.6D2O} of $^7\mathrm{F}_0 \rightarrow {}^5\mathrm{D}_0$ transition with different light polarizations. Dashed: the fitting of the absorption spectrum. 
(\textbf{b}) The fluorescence excitation spectrum. 
(\textbf{c}) The fluorescence lifetime (optical $T_1$) spectrum. 
(\textbf{d}) the restoration of the OD after a spectral hole is burned (hole lifetime).}
\label{fig_spectrum}
\end{figure}

Each \ce{Eu^3+} ion is surrounded by 12 H atoms. 
Each D substitution of an H atom introduces a negative frequency shift to the adjacent \ce{Eu^3+} ion, and increases both fluorescence intensity and excited-state lifetime.
% The \ce{Eu^3+} ions with all 12 D neighbors are on the low frequency side (around -16~GHz), having the strongest fluorescence intensity and longest optical lifetime.
To reduce the multi-phonon non-radiative transitions induced by high-energy phonons from O-H bonds, the laser frequency was restricted to the low-frequency side of the fluorescence excitation peak (Fig.~\ref{fig_spectrum}), ensuring that all 12 nearest-neighbor hydrogen atoms of each \ce{Eu^3+} ion were replaced by deuterium. 

The spectroscopic properties of the  $^7\mathrm{F}_0 \rightarrow {}^5\mathrm{D}_0$ transition of \ce{EuCl3.6D2O} are shown in Fig.~\ref{fig_spectrum}, by using the transition frequency of natural abundance \ce{EuCl3.6H2O} (517148.5~GHz) as a reference.
Absorption spectra with different polarizations are shown in Fig.~\ref{fig_spectrum}(a). 
The spectra exhibit distinct plateaus, revealing strong absorption for light polarized perpendicular to the C2 axis and transparency for the orthogonal polarization, indicating its potential as a promising candidate for polarization-based quantum storage\cite{zhou2012realization}.
The fluorescence excitation spectrum in Fig.~\ref{fig_spectrum}(b) is measured by sweeping the laser frequency while monitoring the fluorescence with wavelengths longer than 600~nm, corresponding to the transition of $^5\mathrm{D}_0 \rightarrow {}^7\mathrm{F}_{n}, (n\geq1)$.
The optical excited state lifetime (fluorescence lifetime) over absorption line is shown in Fig.~\ref{fig_spectrum}(c).
All experiments are performed within the spectral region of -18~GHz to -16~GHz, and the maximum storage efficiency was observed at approximately -16.5~GHz.

The fittings are performed by using similar assumptions in \cite{rose2013JoL}: 
(1) each D induces an identical frequency shift $\Delta\nu$; 
(2) each neighboring H(D) contributes an identical non-radiative decay rate $\Gamma_H$ ($\Gamma_D$); 
(3) D and H are uniformly distributed in the crystal; 
and (4) the absorption line of natural abundance crystals is Gaussian.
Then we have the expression for the OD, fluorescence intensity and lifetime:

%\begin{equation}
\begin{align*}
OD(\nu) =&  k_1 \cdot \sum_{n=0}^{12} p(n) \cdot e^{-\frac{(\nu-n\cdot \Delta\nu)^2}{\sigma^2}}\\
FL(\nu) =& k_2 \cdot \sum_{n=0}^{12} p(n) \cdot r(n) \cdot e^{-\frac{(\nu-n\cdot \Delta\nu)^2}{\sigma^2}}\\
\tau(\nu) = & \frac{\sum_{n=0}^{12}\frac{1}{\Gamma_0+(12-n)\Gamma_1} \cdot e^{-\frac{(\nu-n\cdot \Delta\nu)^2}{\sigma^2}}}{\sum_{n=0}^{12} e^{-\frac{(\nu-n\cdot \Delta\nu)^2}{\sigma^2}}}\\
p(n) =& \frac{12!}{(12-n)!n!} \cdot c^n(1-c)^{12-n}\\
r(n) =& \frac{\Gamma_r}{\Gamma_0 + (12-n)\Gamma_1}\\
\end{align*},
%\end{equation}

where the absorption spectrum $OD(\nu)$ and fluorescence excitation spectrum $FL(\nu)$ are composed of 13 Gaussian curves, with an equal linewidth of $\sigma$ and a lineshift of $n\cdot \Delta \nu$, corresponding to \ce{Eu^3+} ions with adjacent D number of 0~12. 
The intensity of each Gaussian curve is decided by: the proportion of each class $p(n)$ for the OD spectrum, and the ratio of radiative decay rate to total decay rate $r(n)$ for the fluorescence spectrum. 
The observed overall fluorescence lifetime $\tau(\nu)$ is a weighted average of the fluorescence lifetimes of different ion classes, weighted by their atomic proportions. It can be easily verified that the weighted average of lifetime $\tau$ is accurate, rather than the decay rate $1/\tau$.
$\Gamma_1=\Gamma_H-\Gamma_D$ is defined as the difference in the non-radiative transition rates of H and D atoms, and $\Gamma_0 $ represents the total rate of radiative and other non-radiative transitions, excluding those due to D atoms.
The radiative transition rate $\Gamma_r$ cannot be obtained by fitting the experiment data due to its overlap with the arbitrary constant $k$.
The proportion of each class $p(n)$ is given by the binomial distribution with D concentration $c$.

The concentration of D is first fitted by using the absorption spectrum (Fig.~\ref{fig_spectrum}(a)) since their high-frequency edge accurately reflects the proportion of each class $p(n)$, which is determined solely by the concentration parameter.
This fitting gives a D concentration of $c=92.0\%$.
Then the decay rate $\Gamma_{0,1}$, lineshift $\Delta \nu$ and broadening of each class ions $\sigma$ are independently fit to the fluorescence intensity (b) and lifetime (c).
These two fittings give an identical decay rate of $\Gamma_0 = 0.39$~MHz, $\Gamma_1 = 0.56$~MHz. It is close to the result of $\Gamma_0 = 0.38$~MHz, $\Gamma_1 = 0.67$~MHz by Ahlefeldt et al.\cite{rose2013}.
In the three fittings, the lineshift $\Delta \nu$ and linewidth $\sigma$ differ by 10\%, yielding typical values of $\sigma=1.0$~GHz (FWHM = 2.35~GHz), $\Delta \nu=1.33$~GHz.

Fig.~\ref{fig_spectrum}(d) shows the OD restoration of a spectral hole over time.
A three-exponential fit gives the hyperfine lifetime of 1.8 s, 2.2 min, and 27 min, corresponding to different spin relaxation mechanisms \cite{ahlefeldt2016ultranarrow}.
The optical coherence time is measured to be 426~$\mu\mathrm{s}$ and 1060~$\mu\mathrm{s}$ at zero field and a magnetic field of 0.11 T along the C2 axis by the Hahn echo method, respectively.

The 8\% remaining H concentration not only reduces the proportion of \ce{Eu^3+} with 12 neighboring D, but also increases the lattice distortion and thus linewidth. The linewidth of 2.35~GHz is $\sim23$ times that of natural abundance crystal (concentration of 99.985\%, linewidth of 101~MHz) \cite{rose2013}.
If the D concentration rises to 99.5\%, the linewidth of each class ion will reduce to 150~MHz \cite{rose2013}. 
%Then the spectral density has a 15.6-fold enhancement (2.35~GHz / 150~MHz) and the concentration of Er$^{3+}$ with 12 D neighbor ($p(12)$) has a 2.56-fold increase. 
%Applying this 40-fold enhancement in total to the OD of $\sim20$, the OD will up to 800,  
Applying these concentration and bandwidth, our fitting gives a maximum OD of 66. The absorption will be up to 330 cm$^{-1}$ considering the 2 mm thickness of the current crystal.

\section{Analytical model for echo dynamics}
\label{analytical}

To interpret the physical mechanisms underlying the AFC and slow-light protocols, specifically the interplay between absorption and dispersion, we develop both an analytical framework to interpret the echo dynamics and a numerical method for precise experimental fitting.

In the weak-field limit the field propagation equation in the frequency domain can be written as (according to the Eq.~(4-6) in \cite{Spectral_hole}):

\begin{equation}\label{e1}
     \partial_z \tilde{E}(z,\omega) + \frac{i\omega}{c}\tilde{E}(z,\omega) = -\frac{\tilde{\alpha}(\omega)}{2}\tilde{E}(z,\omega),
\end{equation}
where the complex absorption coefficient $\tilde{\alpha}(\omega)$ can be written in the convolution (denoted as '$\ast$') form 
\begin{align}   
    \tilde{\alpha}(\omega) &= \frac{i\alpha_M}{\pi}\cdot g(\omega)\ * \frac{1}{\omega-i \gamma_{ab}},\label{e2}\\
     \alpha_M &= \pi G_0 \frac{k \mu_{ab}^2}{\hbar \epsilon_0},
\end{align}
where $\gamma_{ab}$ and $\mu_{ab}$ are the homogeneous linewidth and dipole moment of transitions between the ground and excited state, respectively. $G_0$ and $\alpha_M$ are the uniform inhomogeneous distribution and absorption coefficient before hole burning, respectively. $g(\omega)=\frac{G(\omega)}{G_0}$ is a normalized inhomogeneous distribution, $k$ is the wavenumber.

By using the convolution theorem and the Fourier transform relation of 
\begin{align}
    &\mathcal{F}[\tilde{\alpha}(\omega)] = \frac{i\alpha_M}{\pi}\cdot \mathcal{F}[g(\omega)]\cdot \mathcal{F}[\frac{1}{\omega-i \gamma_{ab}}], \\
     &\mathcal{F}[\frac{1}{\omega-i \gamma_{ab}}] = -i\sqrt{2\pi}\cdot e^{-\gamma_{ab}t}\cdot u(t) ,
\end{align}
one can transform the propagation equation back to the time domain
\begin{align}\label{e6}
     \partial_z E(z,t) + \frac{1}{c}\partial_t{E}(z,t) = -\frac{\alpha_M}{\sqrt{2\pi}}(\mathcal{F}[g(\omega)]\cdot e^{-\gamma_{ab}t}\cdot u(t))*{E}(z,t),
\end{align}
where $u(t)$ is the Heaviside step function ($t<0, u=0; t=0, u=1/2; t>0, u=1$).
Because the coherence time is much longer than the storage time, $e^{-\gamma_{ab}t}$ term can be ignored.
The second term on the left-hand side of Eqs. (\ref{e1}) and (\ref{e6}), a time derivative term, describes only a trivial transit time of $L/c$ for light traversing the crystal of length $L$, and can therefore be disregarded. 
This assumption is also adopted in Eq.~(10) of \cite{Spectral_hole} and Eq.~(A11) of \cite{afzelius2009multimode}.
Then the simplified evolution equation (no longer a propagation equation) is given by
\begin{align}\label{e7}
     \partial_z E(z,t)= -\frac{\alpha_M}{\sqrt{2\pi}} (\mathcal{F}[g(\omega)]\cdot u(t))*E(z,t).
\end{align}

To obtain an analytical solution, the AFC is treated as a periodically extended spectrum of infinite width. Under this assumption, the Fourier transform of the absorption spectrum yields a series of Dirac delta functions, with coefficients $\alpha_n$ determined by the Fourier series expansion of the comb shape \cite{2010Efficiency}:
\begin{align}\label{e8}
\alpha_M \mathcal{F}[g(\omega)] &= \sqrt{2\pi}\cdot\sum_{n=0}^{\infty} \alpha_n \cdot \delta(t-nT) ,\\
    \alpha_M g(\omega) &= \sum_n \alpha_n e^{-in\omega T}.
\end{align}

For the square AFC with angular frequency period of $2\pi/T$ and Finesse of $F$, the coefficient will be 
 \begin{align}
     \alpha_0 &= \alpha_M/F,\\
     \alpha_n &=  \alpha_0\cdot sin(\frac{n \pi}{F})/\frac{n \pi }{F}.
 \end{align}

Because the pulse duration is much shorter than the memory time, and considering $f(t) \ast \delta(t-T) = f(t-T)$, the field can be written as a series of temporally resolvable pulses $E(z,t) = \sum_{n=0}^{+\infty} a_n(z)E(t-nT)$, where $a_n(z)$ is the amplitude of each pulse and $E(t)$ is the original input pulse, which is only valid near $t=0$.
The simplified evolution equation becomes 
\begin{equation}\label{e12}
    \sum_{n=0}^{\infty}\partial_z a_n(z) E(t-nT) = 
-\sum_{n=0}^{\infty}\sum_{m=-\infty}^{\infty} a_n(z) \cdot\alpha_m \cdot u(m)\cdot E(t-nT-mT).
\end{equation}
It describes how the $n$th echo is driven by all the other echoes. 
In which the step function $u(m)$ reveals the causality, i.e. the later pulse is only influenced by earlier pulses or itself. 

One can then extract the terms which correspond to the $n$th Echo $E(t-nT)$
\begin{equation}\label{e13}
    \begin{split}   
    \partial_z a_0(z) &= -\tfrac{1}{2}\alpha_0 a_0(z) ,\\
    \partial_z a_1(z) &= -\tfrac{1}{2}\alpha_0 a_1(z) - \alpha_{1}a_0(z),\\
    \partial_z a_n(z) &= -\tfrac{1}{2}\alpha_0 a_n(z) - \alpha_{1}a_{n-1}(z) - ... - \alpha_{n}a_0(z).
    \end{split}
\end{equation}

Here are the solutions of the first 6 echo amplitudes
\begin{equation}
    \label{e15}
    \begin{split}
    a_0(z) &= e^{-\tfrac{1}{2}\alpha_M z/F},\\
    a_1(z) &= a_0(z)\cdot[-z\cdot\alpha_{1}],\\
    a_2(z) &= a_0(z)\cdot[\frac{z^2}{2}\cdot\alpha_{1}^2-z\cdot\alpha_{2}],\\
    a_3(z) &= a_0(z)\cdot[-\frac{z^3}{3!}\cdot\alpha_{1}^3+2\frac{z^2}{2!}\cdot\alpha_{1}\alpha_{2}-z\cdot\alpha_{3}],\\
    a_4(z) &= a_0(z)\cdot[\frac{z^4}{4!}\cdot\alpha_{1}^4-3\frac{z^3}{3!}\cdot\alpha_{1}^2\alpha_{2}+2\frac{z^2}{2!}\cdot\alpha_{1}\alpha_{3}+\frac{z^2}{2!}\cdot\alpha_{2}^2-z\cdot\alpha_{4}],\\
    a_5(z) &= a_0(z)\cdot[-\frac{z^5}{5!}\cdot\alpha_{1}^5 +4\frac{z^4}{4!}\cdot\alpha_{1}^3\alpha_{2} -3\frac{z^3}{3!}\cdot\alpha_{1}^2\alpha_{3}  \\
    &\quad-3\frac{z^3}{3!}\cdot\alpha_{1}\alpha_{2}^2 +2\frac{z^2}{2!}\cdot\alpha_{1}\alpha_{4}+2\frac{z^2}{2!}\cdot\alpha_{2}\alpha_{3} - z\cdot\alpha_{5}].
    \end{split}
\end{equation}
The efficiency of each echo is given by $|a_n(L)|^2$. The first two expressions are given by Bonarota et al. \cite{2010Efficiency}. 

\section{Numerical calculation for precise output field}
\label{numerical}
To obtain the theoretical solution of the output field in the time domain $E(L,t)$, we swap the order of the Fourier transform and integration compared to the analytical method:

\begin{align}
    \partial_z \tilde{E}(z,\omega) &= -\frac{\tilde\alpha(\omega)}{2}\tilde{E}(z,\omega),\\
    E(L,t) &=\mathcal F[\tilde{E}(L,\omega)] = \mathcal F[ \tilde{E}(0,\omega)\cdot e^{-\tilde\alpha(\omega)L/{2}} ] \label{e17}
\end{align}

An analytical solution is difficult to obtain directly. Furthermore, direct numerical calculation of $\tilde\alpha(\omega)$ (Eq.~(\ref{e2})) in the exponential part will also cause non-negligible errors. 
Here, we derive the analytical solution of $\tilde\alpha(\omega)$ by assuming a square-shaped atomic absorption distribution $g(\omega)$: there are $n$ rising or falling edges in $g(\omega)$, with frequencies $f_1, f_2, \dots ,f_n$. 
Then the integral of  $\tilde\alpha(\omega)$ in Eq.~(\ref{e2}) is divided into $n+1$ segments: $(-\infty,f_1), (f_1,f_2), \dots , (f_n, \infty)$. Let $g(\omega)=g_{ab}$ when $\omega \in (f_a,f_b)$, this segment integral will be

\begin{align}
    \int_{f_a}^{f_b} \mathrm{d}\delta \, g_{ab}\frac{1}{\omega-i\gamma-\delta}
    &= g_{ab}\int_{f_a}^{f_b} \mathrm{d}\delta\, \frac{\omega-\delta}{(\omega-\delta)^2+\gamma^2} + i g_{ab} \gamma \int_{f_a}^{f_b} \mathrm{d}\delta\, \frac{1}{(\omega-\delta)^2+\gamma^2}\\
    &= -\frac{g_{ab}}{2} \ln[(\delta-\omega)^2+\gamma^2]\bigg|_{\delta=f_a}^{f_b} + i g_{ab} \arctan \left( \frac{\delta-\omega}{\gamma} \right) \bigg|_{\delta=f_a}^{f_b}.
\end{align}

The analytical expression of the square $\tilde\alpha(\omega)$ will be the sum of $n+1$ segment integrals. Due to the atomic decoherence $\gamma$, the absorption spectrum (i.e., the real part of $\tilde\alpha(\omega)$) is no longer strictly square. The time domain output field can be obtained by applying the Fast Fourier Transform in Eq.~(\ref{e17}).
This method is also applicable to all storage schemes with a static spectrum, including AFC and slow-light storage.

\bibliography{samp}

\end{document}